\documentclass{aa}  
\usepackage{graphicx}
\usepackage{txfonts}
\usepackage{natbib}
\bibpunct{(}{)}{;}{a}{}{,} 

\newcommand{\xmm}{{\it XMM-Newton}}

\newcommand{\bona}{{\sl bona fide}}
\newcommand{\xest}{{\it XMM-Newton Extended Survey of the Taurus
Molecular Cloud}}

\begin{document}
   \title{A U-band survey of brown dwarfs in the Taurus Molecular
Cloud with the XMM-Newton Optical/UV Monitor}

\titlerunning{A U-band survey of BDs in the TMC with the OM}


   \author{N.\ Grosso\inst{1}
\and M.\ Audard\inst{2}
\and J.\ Bouvier\inst{1}
\and K.~R.\ Briggs\inst{3}
\and M.\ G{\"u}del\inst{3}}

   \offprints{N.\ Grosso}

   \institute{Laboratoire d'Astrophysique de Grenoble,
              Universit{\'e} Joseph-Fourier,
              F-38041 Grenoble cedex 9, France\\
              \email{\tt Nicolas.Grosso@obs.ujf-grenoble.fr}
         \and Columbia Astrophysics Laboratory, Columbia University,
              550 West 120th Street, New York, NY 10027, USA
         \and Paul Scherrer Institut, 5232 Villigen und W{\"u}renlingen,
                Switzerland
             }

   \date{Received 12 April 2006 / Accepted 16 June 2006}


  \abstract
   {}
   {We aim to characterize the $U$-band variability of young
brown dwarfs in the Taurus Molecular Cloud and discuss its origin.}
   {We used the \xest, where a sample of 11
     young \bona~brown dwarfs (spectral type later than M6)
     were observed simultaneously in X-rays with \xmm~and in the
     $U$-band with the \xmm~Optical/UV Monitor (OM).}
   {We obtained upper limits to the $U$-band emission of 10 brown dwarfs
($U$$\ge$19.6--20.6\,mag), whereas 2MASS\,J04141188+2811535 was
detected in the $U$-band. Remarkably, the magnitude of this brown
dwarf increased regularly from $U$$\sim$19.5\,mag at the beginning of
the observation, peaked 6\,h later at
$U$$\sim$18.4\,mag, and then decreased to
$U$$\sim$18.65\,mag in the next 2\,h. The first OM $U$-band
measurement is consistent with the quiescent level observed about one
year later thanks to ground follow-up observations. This brown dwarf was not
detected in X-rays by \xmm~during the OM observation.}
   {We discuss the possible sources of $U$-band variability for
this young brown dwarf, namely a magnetic flare, non-steady accretion onto
the substellar surface, and rotational modulation of a hot spot. 
We conclude that this event is related to accretion from
a circumsubstellar disk, where the mass accretion rate was about
a factor of 3 higher than during the quiescent level.}
   \keywords{Stars: low-mass, brown dwarfs -- Stars: individual:
2MASS\,J04141188+2811535 -- X-rays: stars -- ISM:
individual objects: the Taurus Molecular Cloud
               }

   \maketitle
%

\section{Introduction}

Young brown dwarfs (BDs) share similar properties with
pre-main-sequence low-mass stars, suggesting that they are also
undergoing a T~Tauri phase. BDs have accretion disks
\citep[e.g.,][]{jayawardhana03,natta04,mohanty05}, and photometric
\citep{scholz04a,scholz04b}, as well as spectroscopic variability
\citep{scholz05a,scholz06}, pointing to variable mass
accretion. Besides, BDs also have an active corona, emitting soft X-rays
\citep[e.g.,][]{neuhaeuser99,preibisch05}. 

The gas accreting from the circumstellar disk onto the stellar surface
of T~Tauri stars produces a UV/blue excess emission \citep[see review
by][]{bouvier06}. The $U$-band luminosity of T~Tauri stars is directly
related to the accretion luminosity \citep{gullbring98}. However, due
to their faintness in the optical, BDs are poorly studied at short
wavelengths \citep[][]{gizis05}.
During the \xest~\citep[XEST;][]{guedel06b},
the X-ray emission of a sample of 16 \bona~BDs 
(spectral type later than M6) was
surveyed with \xmm~\citep{grosso06}. Eleven of these BDs were
located less than $\sim$$8.5\arcmin$--\,$\sim$12$\arcmin$ away from the
prime target of 9 \xmm~pointings, which allowed for simultaneous observations with the
\xmm~Optical/UV Monitor \citep[OM;][]{mason01}, that we report here.

We present the constraints on the $U$-band photometry of these BDs
obtained with the OM in Sect.~\ref{aperture}. In
Sect.~\ref{variability}, we investigate the variability of
2MASS\,J04141188+2811535 (shortened to 2MASS\,J0414 henceforth) in the
$U$-band and in X-rays. We discuss the origin of the $U$-band event
observed from this BD in Sect.~\ref{discussion}.

\begin{table*}[!ht]
\caption{The sample of the 11 young BDs, with spectral type later
    than M6, surveyed with the
 \xmm~Optical/UV Monitor (OM) in the XEST \citep{guedel06b}. Columns~(1)
and (2) give the BD names and the 2MASS counterparts, respectively. 
Column~(4) is the references of the spectral types in Col.~(3), with
B02=\citet{briceno02}, G06=\citet{guieu06}, L06=\citet{luhman06}, and M05=\citet{muzerolle05b}.
Column~(5) gives the disk accretion rate when available
  (\citealt{mohanty05}; for 2MASS\,J0414, \citealt{muzerolle05b}).
Column~(6) is the XEST field number, plus the source
number if there is an X-ray detection \citep{grosso06}. 
Columns~(7) and (8) indicate the duration of one single full-frame/imaging window
exposure and the total number of similar exposure obtained on the BD. 
Column~(9) gives the $U$-band magnitude or the
$U$-band limiting magnitude of each window exposure for a 3$\sigma$
point-source detection (this work). For 2MASS\,J0422, the limiting
magnitude was computed for the broader $UVW1$ filter (200--400\,nm),
which includes the $U$-band.}
\label{table:bd}
\vspace{-0.25cm}
\centering
\begin{tabular}{lclcclccr}
\hline\hline
\noalign{\smallskip}
\multicolumn{1}{c}{Name} & 2MASS & 
\multicolumn{1}{c}{SpTyp} &   \multicolumn{1}{c}{Ref.} & $
\log{\dot{M}}$ & \multicolumn{1}{c}{XEST} & \multicolumn{1}{c}{Window
exposure} & $N$ & \multicolumn{1}{c}{$U$}\\
     &                &                  &    & M$_\odot$\,yr$^{-1}$ &
& \multicolumn{1}{c}{ks} & & \multicolumn{1}{c}{mag} \\
\multicolumn{1}{c}{(1)} & (2) & \multicolumn{1}{c}{(3)} &
\multicolumn{1}{c}{(4)} & \multicolumn{1}{c}{(5)} & \multicolumn{1}{c}{(6)} &
\multicolumn{1}{c}{(7)} & \multicolumn{1}{c}{(8)}& \multicolumn{1}{c}{(9)}\\
\hline
2MASS\,J0414  & J04141188+2811535   & M6.25 & M05  & -10      & 20     & 2.5 & 10  & 19.5\\   
2MASS\,J0421  & J04215450+2652315   & M8.5 & L06  & \dotfill & 11     & 5.0 & 5.0 & $\ge$20.4\\
2MASS\,J0422  & J04221332+1934392   & M8   & L06  & \dotfill & 01-062 & 3.5 & 4.0 & $\ge$19.9\\
KPNO-Tau\,4   & J04272799+2612052   & M9.5  & G06  & -11.1    & 02     & 5.0 & 4.6 & $\ge$20.4 \\ 
KPNO-Tau\,5   & J04294568+2630468   & M7.5  & B02  & n.a.     & 15-044 & 1.3 & 3.9 & $\ge$19.8 \\ 
KPNO-Tau\,6   & J04300724+2608207   & M9    & G06  & -10.8    & 14     & 5.0 & 2.8 & $\ge$20.6 \\ 
KPNO-Tau\,7   & J04305718+2556394   & M8.25 & B02  & -11      & 14     & 5.0 & 2.8 & $\ge$20.6 \\ 
CFHT-Tau\,5   & J04325026+2422115   & M7.5  & G06  & n.a.     & 04-003 & 1.5 & 4   & $\ge$19.8 \\ 
KPNO-Tau\,9   & J04355143+2249119   & M8.5  & B02  & n.a.     & 08     & 1.6 & 4   & $\ge$19.6 \\ 
CFHT-BD-Tau\,2& J04361038+2259560   & M7.5  & B02  & n.a.     & 08     & 1.3 & 8   & $\ge$19.7 \\ 
CFHT-Tau\,6   & J04390396+2544264   & M7.25 & G06  & \dotfill & 05-005 & 5.0 & 3.7 & $\ge$20.5 \\ 
\hline
\end{tabular}
\end{table*}

\section{The OM observations of BDs in the Taurus Molecular Cloud}
\label{aperture}

Table~\ref{table:bd} gives the list of the 11 BDs of the Taurus
  Molecular Cloud (TMC) surveyed with the OM in
9 \xmm~pointings. About half of these \xmm~pointings -- namely
XEST-02, 05, 11, 14 -- were made with the OM in full-frame
imaging mode, which provides an image of the whole OM field of view
(17\arcmin$\times$17\arcmin) with $2\times2$ binned (low) spatial
resolution ($1\arcsec\times1\arcsec$). The other \xmm~pointings --
namely XEST-01, 04, 08, 15, 20 -- were made with the OM in
the default imaging mode, which consists of a sequence of 5 exposures
with 5 imaging windows -- allowing us to cover 92\% of the total
(17\arcmin$\times$17\arcmin) OM field-of-view -- plus a small central
imaging window (1.7\arcmin$\times$1.7\arcmin). In each of the 5
exposures, one of the 5 large imaging windows covers a large fraction
of the field-of-view with $1\arcsec\times1\arcsec$ spatial resolution,
while the small imaging window ensures continuous monitoring of the
prime target at the center of the field-of-view with
$0\farcs5\times0\farcs5$ spatial resolution. We indicate in Col.~(7)
of Table~\ref{table:bd} the duration of one single full-frame/imaging window
exposure and in Col.~(8) the total number of similar exposures obtained on the BD.

In a nutshell the OM detector is a micro-channel plate (MCP)
intensified CCD detector operated as a photon-counting instrument.
First, an optical/UV photon collected by the 30\,cm-diameter
telescope mirrors hits a photo-cathode located at the backside of the
detector entrance window. The photoelectron is then amplified by two
MCPs. The resulting electron avalanche is converted to a photon splash
by a fluorescent phosphorus layer at the top of a fiber taper, which
illuminates the CCD pixels. Finally, the onboard
software determines the accurate detector position of the photon
splash's centers in each readout frame using a centroiding algorithm,
which subsamples each CCD pixels by $8\times8$ in-memory pixels (about
$0\farcs5\times0\farcs5$ on the sky).

We use the PERL script {\tt omichain} of the {\it XMM-Newton Science
Analysis System} (SAS; version 6.5) with default parameters to run
the OM imaging mode pipeline. We corrected OM sky source positions
in the observation source list (*OM*SWSRLI* fits files) of residual (a
few arc seconds) boresight errors by using the 2MASS positions of OM
sources as references. Only one BD was detected with the OM in the 
$U$-band (305--390\,nm, $\lambda_{\rm eff}=344$\,nm). We made a visual
inspection of the stacked sky image covering the full OM field
produced by the pipeline to look for possible faint BD missed by the
OM source detection algorithm in individual imaging window exposure,
but we found no other detection.

For each non-detected BD, we compute, using an {\tt IDL} procedure, the
limiting magnitude of each window exposure for a 3$\sigma$
point-source detection at the detector position of the BD because
this local value is not provided by any {\tt SAS} command. We first need 
to determine the relation between sky coordinate and detector coordinate. We
rotate the fits sky image of the exposure window (*OM*SIMAG* fits
file) using the roll angle of spacecraft pointing, we use details of
window configuration (MODES extension of *OM*IMAG\_* fits file) to
compute absolute position within OM frame, and we apply OM
distortions (introduced mainly by the fiber taper)\footnote{Provided
by the POLYNOM\_MAP2 extension of the current calibration file
OM\_ASTROMET\_0011.CCF.}. Comparison with the observation
source list, which provides both sky and detector source coordinates,
helps to fix small systematic offsets in our coordinate transformation. 
We apply it to BD sky coordinates with the boresight error added to find
the proper BD detector coordinates. 
Inside a $6\arcsec$-radius disk centered at this position we estimate
the sky count rate on the detector using a
2$\sigma$-clipping procedure\footnote{See {\tt IDL DAOPHOT}-type photometry
procedures: {\tt aper.pro} and {\tt meanclip.pro} available at {\tt
http://idlastro.gsfc.nasa.gov}\,.}. We check the absence of bad pixels
in the aperture using the QUALITY extension of this fits file.
We compute the corresponding count
rate error at the 3$\sigma$ level assuming a Poisson noise. We derive 
the count rate error of incident photons taking into account coincidence-loss,
deadtime, and temporal sensitivity degradation correction (see
Appendix~\ref{appendix}). Finally using the OM zero point
in the $U$-band filter ($ZPTU=18.26$\,mag), we convert this incident
count rate to $U$-band magnitude. Table~\ref{table:bd} gives the upper
limits in the $U$-band. We conclude that the 10 non-detected BDs have
$U$$\ge$19.6--20.6\,mag.

\begin{figure}[!t]
\centering
\includegraphics[width=0.8\columnwidth]{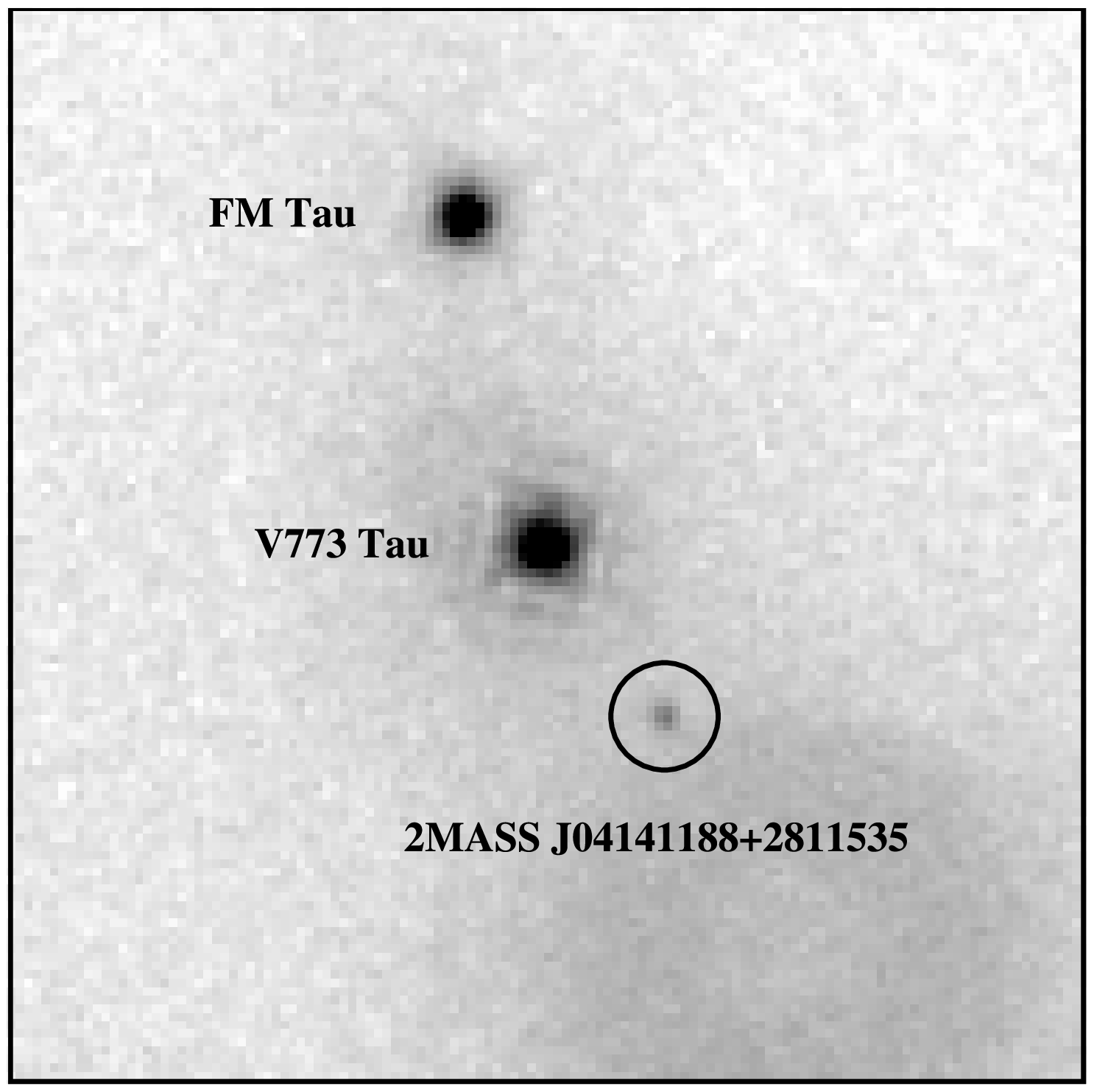}
 \caption{The \xmm~Optical/UV Monitor (OM) image in the $U$-band of
the young brown dwarf (BD) 2MASS\,J0414. This sky image is the sum of ten OM
central window exposures; the total on-source exposure was about 7\,hours.
The image size is $2\arcmin \times 2\arcmin$ with North up and East
left, with pixels of $1\arcsec \times 1\arcsec$; the stretch of the color
scale is logarithmic. This image is centered on the weak-line T~Tauri
star V773\,Tau located only $24\arcsec$ away from the BD. 
The Classical T~Tauri star FM\,Tau is also visible. The
$6\arcsec$-radius circle centered on 2MASS\,J0414 indicates the
aperture used in the photometry measurement. The FWHM of the OM in the
$U$-band is $\sim$$1\farcs6$.
}
\label{fig:om}
\end{figure}

This sample of 10 non-detected BDs is composed nearly equally of accreting
and non-accreting sources \citep[see][]{grosso06}. The former
have $U$$\ge$20.4\,mag, and an average visible extinction of 1.5\,mag,
which leads, after correction of the extinction \citep[][ with $R_{\rm
V}$=3.1]{cardelli89}, to $U$$\ge$19.4\,mag and the corresponding
$U$-band luminosity\footnote{ We adopt a distance of 140\,pc for the
TMC and use the conversion factor from the OM
count rates to fluxes for M-type stars available at {\tt
http://xmm.vilspa.esa.es/sas/documentation/watchout/\-uvflux.shtml}\,.},
$L_{\rm U} \la 5.4\times 10^{29}$\,erg\,s$^{-1}$. Assuming that the
$U$-band luminosity is directly related to the accretion luminosity as
observed for the T~Tauri stars \citep{gullbring98}, we can derive an
estimate of the upper limit of the disk accretion rate. Taking typical
values for the BD mass ($\sim$0.07\,M$_\odot$), substellar radius ($R_\star \sim
0.5$\,R$_\odot$), and disk inner radius ($\sim$4.5\,R$_\star$) leads
to $\dot{M} \la 1 \times 10^{-10}$\,M$_\odot$\,yr$^{-1}$, which is consistent
with the disk accretion rates obtained from spectroscopy for some of these
BDs (Table~\ref{table:bd}). We focus on the only BD detected in the
$U$-band in the following sections.

\section{The variability of 2MASS\,J04141188+2811535}
\label{variability}

Figure~\ref{fig:om} shows the OM image of 2MASS\,J0414, the young BD
detected in the $U$-band with the OM.
This BD is located only $24\arcsec$ away from the weak-line T~Tauri
star V773\,Tau, the prime target of this \xmm~observation located
on-axis\footnote{A limitation in the onboard pixel subsampling
algorithm introduces a modulo 8 pattern, which is removed by the data
reduction pipeline, but residual modulo 8 background fluctuations are
still visible around this bright source.}.
Internal reflection of the light of V773\,Tau ($U$=13.1\,mag) within the
detector entrance window produced an out-of-focus ghost image, a ``smoke ring''
centered $\sim$$40\arcsec$ South-West away from V773\,Tau. 
The default $7\arcsec$--$12.5\arcsec$ radii annulus centered on the
source used in {\tt omichain} to estimate the sky photometry overlaps
the smoke ring. Therefore, the estimate of the sky level is
overestimated, and hence the photometry of 2MASS\,J0414 provided by this
{\tt SAS} task is underestimated. The $6\arcsec$-radius disk aperture used
to estimate the source+sky photometry also partially overlaps the
border of the smoke ring. The default aperture photometry needs to be
refined.\footnote{The OM light curve of 2MASS\,J0414 presented previously
in conference proceedings \citep{grosso06c} was obtained with {\tt
SAS} version 6.1 and default aperture photometry, and was then
strongly affected by the smoke ring.}

\begin{figure}[!t]
\centering
\includegraphics[width=\columnwidth]{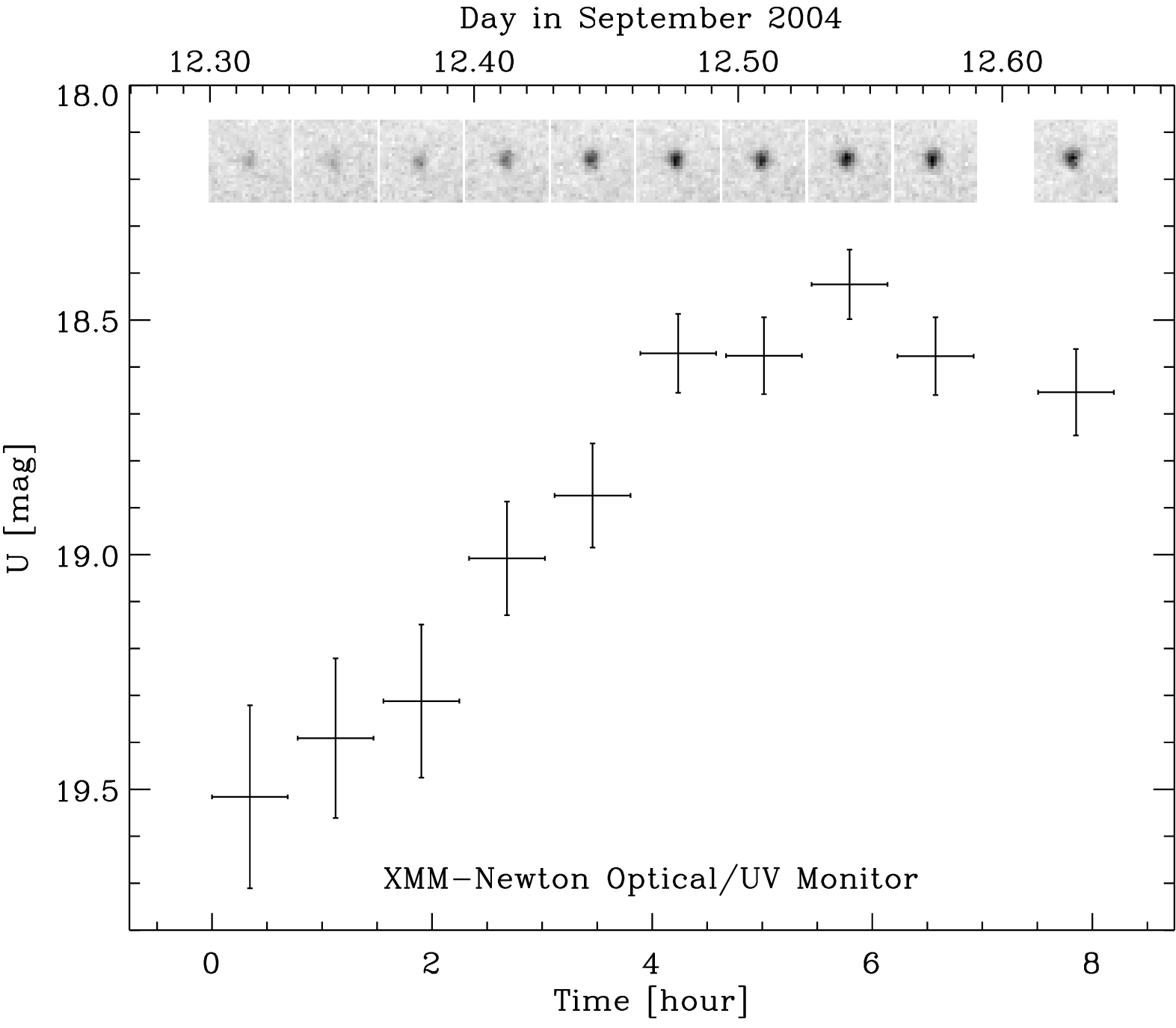}
 \caption{The OM light curve of 2MASS\,J0414 in the $U$-band. The
   image sequence shows enlargements (field-of-view of $12\arcsec
   \times 12\arcsec$
with North up and East left, and with $0.5\arcsec \times 0.5\arcsec$
pixel-size) of the OM central small window exposures (2480\,s) in the
$U$-band; the stretch of the color scale is linear. The OM observed for
$\sim$6\,hours showed an increase in the brightness ($\Delta U\sim-1$\,mag).}
\label{fig:lc}
\end{figure}

We take the sky region on the detector image (here the small central
imaging window; *IMAG\_0000*) as close as possible to the
source. Indeed, the FWHM is about $1\farcs6$ in the $U$-band, and the
PSF wings are limited for this weak source, consequently there is no
obvious signal from the source beyond $4\arcsec$. Therefore, we tune the
sky annulus from $4\arcsec$--$6\arcsec$ pixels, i.e., as close as
possible to the source center. This operation is not allowed in the
{\tt SAS} command {\tt omsource}, and we had to use our own {\tt IDL}
procedure for this aperture photometry. We keep the canonical
$6\arcsec$-radius for the source+sky aperture because this is the
size that was used in the calibration of the
coincidence-loss. Moreover, taking a smaller radius would imply the
use of the PSF calibration (note that the PSF shape is flux dependent due to
the non-linearity of the detector), and hence the introduction of other
calibration uncertainties. We measure the count rates of sky+source
and sky apertures on the detector, we correct the sky count rate from
the difference in aperture area, and we apply non-linearity corrections
(Appendix~\ref{appendix}) to both detector count rates to retrieve the
count rate of sky+source and source incident photons. The count rate
of source incident photons is simply the difference between these two
numbers. 

\begin{table}[!t]
\caption{Optical CCD photometry of 2MASS\,J0414
obtained with the 1.3-m McGraw-Hill telescope at the MDM
observatory on October 25, 2005. Filter values are the average of two
observations obtained during the same night (05$^{\rm h}$19$^{\rm
m}$--05$^{\rm h}$34$^{\rm m}$ UT and 11$^{\rm h}$46$^{\rm
m}$--11$^{\rm h}$55$^{\rm m}$ UT).
}
\label{table:mdm}
\centering
\begin{tabular}{cccc}
\hline\hline
Name  &  $U$           &  $B$           & $V$ \\
& mag & mag & mag\\
\hline
2MASS\,J0414 & 19.49$\pm$0.25 & 19.13$\pm$0.10 & 19.56$\pm$0.07  \\
\hline
\end{tabular}
\end{table}

\begin{figure}[!ht]
\centering
\includegraphics[width=0.8\columnwidth]{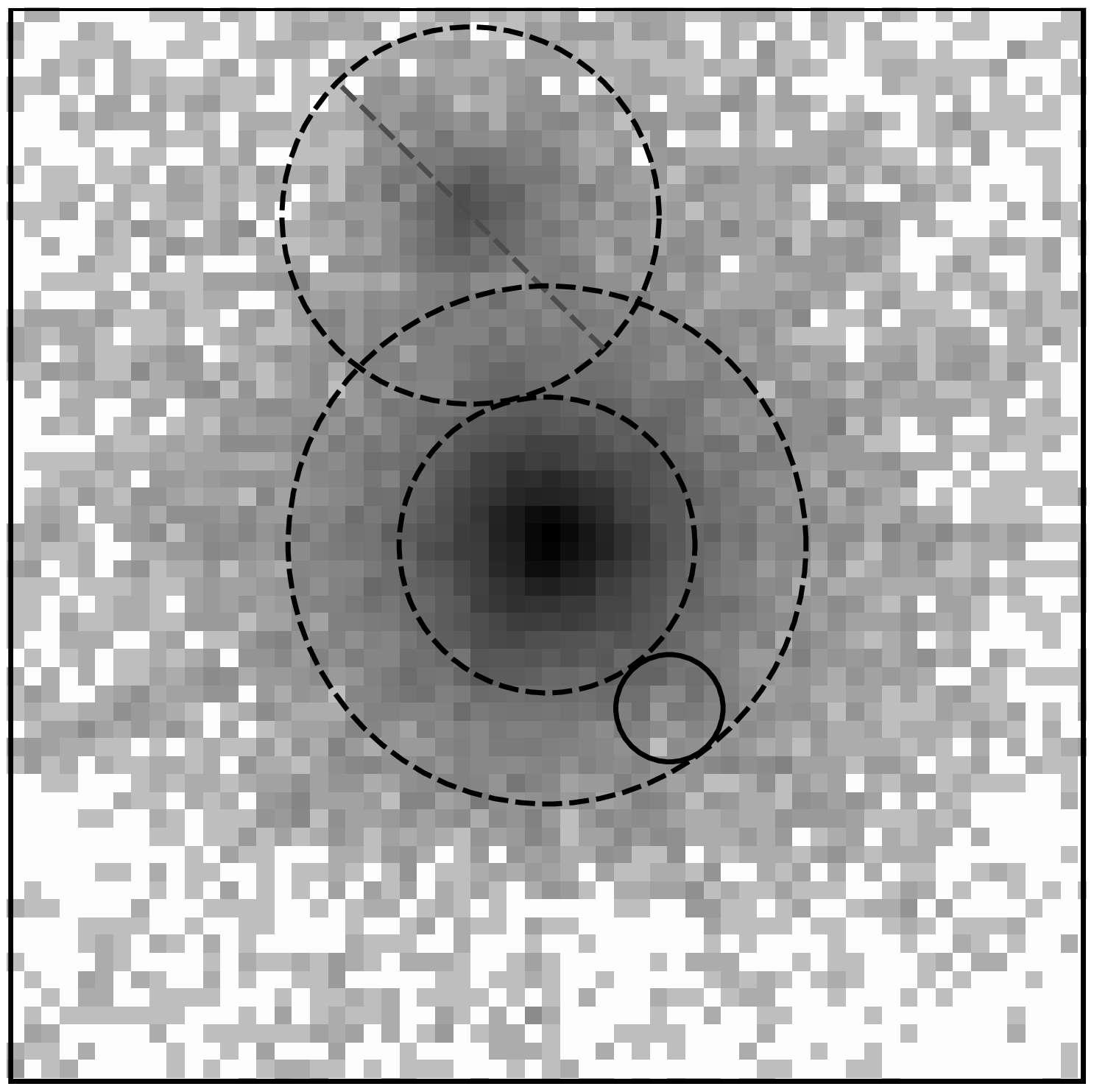}
 \caption{X-ray image of V773\,Tau in the 0.5--2\,keV energy band
obtained with \xmm/EPIC (pn+MOS1+MOS2). The field-of-view and
orientation are the same as that in Fig.~\ref{fig:om}. The stretch of the
color scale is logarithmic. 2MASS\,J0414 is located on the PSF wings of V773\,Tau,
where no X-ray counterpart was found with the source detection
algorithm. The  $6\arcsec$-radius circle indicates the region used for
the selection of the background+source X-ray events. The
$12\arcsec$-width annulus centered on V773\,Tau indicates the region
used for the selection of the background X-ray events; the circular region
centered on FM\,Tau, another X-ray source, was excluded from this annulus.
}
\label{fig:x}
\centering
\includegraphics[width=\columnwidth]{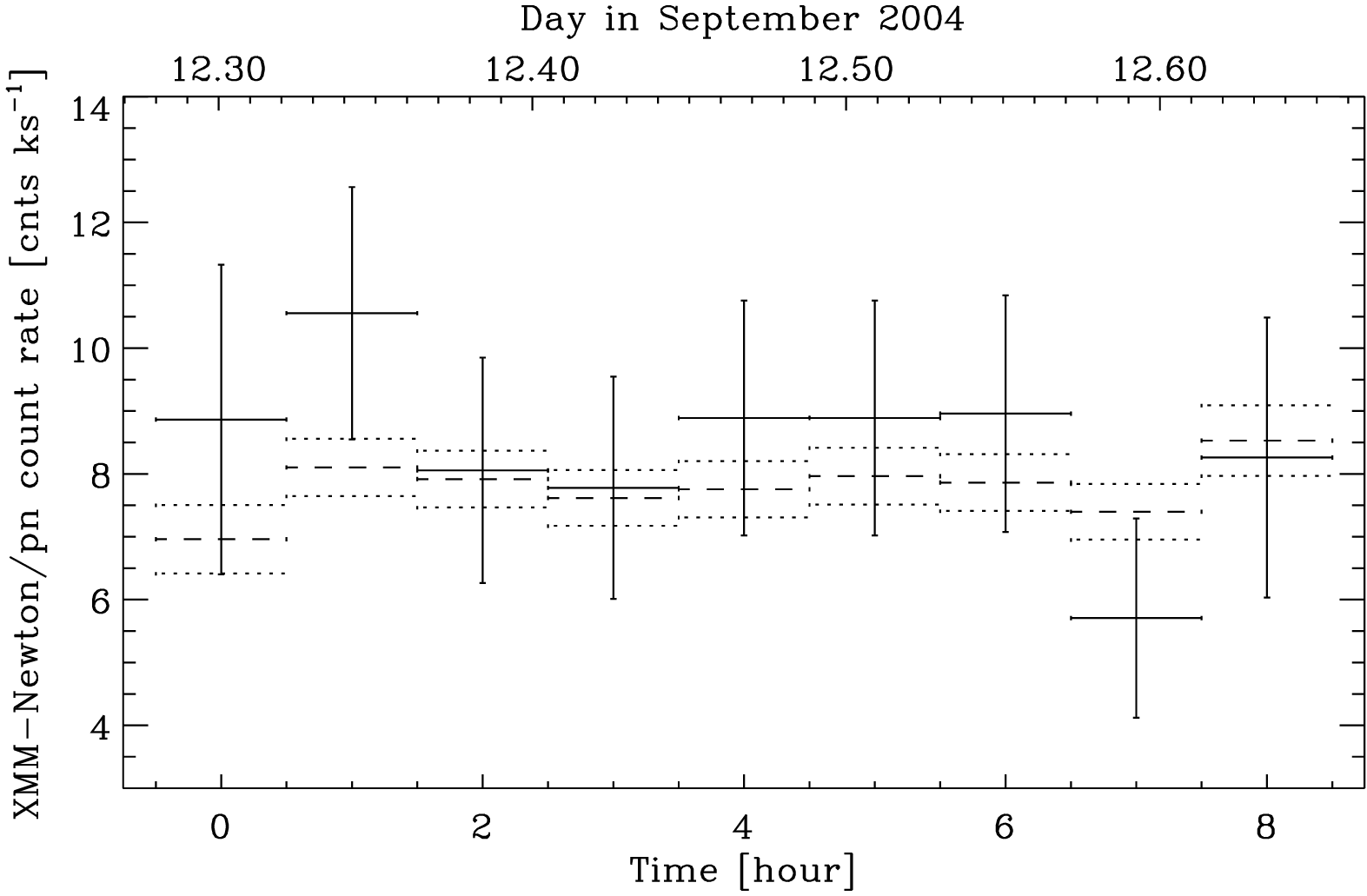}
 \caption{\xmm/EPIC pn X-ray light curve of 2MASS\,J0414. The solid
lines show the source+background light curve at the location of
2MASS\,J0414 (see the definition of selected regions in
Fig.~\ref{fig:x}). The dashed/dotted lines show the estimate of the
background level inside the extraction region for comparison. The time
bin size is one hour. The scale of the time axis is identical to the
one used in Fig.~\ref{fig:lc}. 2MASS\,J0414 is not detected in X-rays.
}
\label{fig:pn}
\end{figure}

The resulting photometry of 2MASS\,J0414 is shown in
Fig.~\ref{fig:lc}. The location of this BD on the small central imaging
window of the OM provided a nearly continuous coverage during $\sim$$8$\,h with
42\,min-exposures. The magnitude of the BD increased regularly from
$U$$\sim$19.5\,mag at the beginning of the observation, peaked 6\,h
later at $U$$\sim$18.4\,mag, and then decreased to $U$$\sim$18.65\,mag
in the next 2\,h.

We obtained, about one year after the OM observation, a follow-up
observation of 2MASS\,J0414 with the 1.3-m McGraw-Hill telescope at the MDM
observatory. This source was observed twice, 6.4\,h apart on October
25, 2005. The source had a constant magnitude of $U=19.49\pm0.25$\,mag
(Table~\ref{table:mdm}). The first OM $U$-band measurement is fully
consistent with this ground photometry; therefore we consider this
$U$-band level as the quiescent level of 2MASS\,J0414.

Figure~\ref{fig:x} shows the simultaneous \xmm/EPIC (pn+MOS1+MOS2) image of
V773\,Tau. 2MASS\,J0414 is located on the highly structured PSF wings
of this bright X-ray source. No X-ray source was detected at this
location with the {\tt SAS} source detection algorithm. The upper
limit on the X-ray luminosity in the 0.5--8\,keV energy range is
$L_{\rm X} \le 1.6\times10^{28}$\,erg\,s$^{-1}$, and
$\log(L_{\rm X}/L_{\rm bol}) \le -3.6$ \citep{grosso06}. For
comparison, BDs detected in X-rays in the TMC have
$\log(L_{\rm X}/L_{\rm bol}) \sim -4.5$ to $-3.0$
\citep{neuhaeuser99,mokler02,grosso06};
therefore this upper limit is not so strong. We also look here for
possible X-ray variability, which could help to detect this BD during
a limited time interval. Figure~\ref{fig:pn} shows the source+background
X-ray light curve for \xmm/EPIC~pn at the position of 2MASS\,J0414,
compared with the estimate of the background level inside the
$6\arcsec$-radius extraction region. No X-ray excess is found above
the background, hence 2MASS\,J0414 is definitively not detected in
X-rays.

\section{Discussion}
\label{discussion}

2MASS\,J0414 was identified as a (young) BD of the TMC by
\citet{luhman04}. It has a visual extinction of 1.1\,mag; an
M6.25 spectral type, which corresponds to an effective
temperature of 2960\,K; and a bolometric
luminosity of 0.015\,L$_\odot$ \citep{luhman04}. From these values the
BD radius can be estimated to 0.5\,R$_\odot$. Using the evolutionary
tracks from \citet{baraffe98},  the BD mass is 0.07\,M$_\odot$. The
only published optical spectrum of 2MASS\,J0414,
obtained on December 15, 2003, exhibits a strong excess of continuum at
short wavelengths plus a strong H$\alpha$ emission \citep[$EW({\rm
H}\alpha)$=250\,\AA;][]{muzerolle05b} indicating accretion from a
circumsubstellar disk \citep{luhman04}, with an estimated accretion
rate of $\sim$$10^{-10}$\,M$_\odot$\,yr$^{-1}$
\citep{muzerolle05b}. Blueshifted absorption components superimposed
on the H$\alpha$ accretion emission line provide evidence for an
outflow \citep{muzerolle05b}.

2MASS\,J0414 has $B-V \sim -0.4$ (Table~\ref{table:mdm}), which
is extremely blue for an object with an M6.25
spectral type. This strong blue excess indicates that this BD is
actively accreting. Following the method used in Sect.~\ref{aperture},
we find that its quiescent level, $U=19.5\pm0.2$\,mag, corresponds to
$\dot{M}=1.9\pm0.4 \times 10^{-10}$\,M$_\odot$\,yr$^{-1}$, which is
consistent with the value of \citet{muzerolle05b} derived from
spectroscopy. Therefore, the high disk accretion rate of 2MASS\,J0414
is likely the reason why we detected only this BD in the $U$-band.

We consider 3 possible sources of $U$-band variability for this young
BD, namely: a magnetic flare, non-steady accretion onto the substellar
surface, and rotational modulation by a hot spot. We discuss these 3
possibilities in turn.

\xmm~observations of coronal flares from active stars have shown an
extremely good temporal correlation between UV and X-ray variations
\citep{mitra05}. A similar result was found in the optical $U$-band
and X-rays in an \xmm~observation of Proxima Cen, where several X-ray
weak flares were preceded by an optical burst \citep{guedel04}. 
We can predict the flare luminosity increase in
the X-rays with the UV-X-ray relationship in flares of
\citet{mitra05}. The increase from quiescent to flare peak in the
$U$-band is equivalent to a spectral luminosity density of
$2.4\times10^{26}$\,erg\,s$^{-1}$\,$\AA^{-1}$. The \citet{mitra05}
relationship, based on the $UVW1$ OM filter \citep{ehle05}, yields
$\sim$$1\times 10^{29}$\,erg\,s$^{-1}$
in the 0.5--8\,keV energy range, which is a
lower limit on the X-ray luminosity increase because the $U$-band
filter covers roughly half of the wavelength range of the $UVW1$ filter
(200--400\,nm). Therefore, a UV flare would produce an X-ray flare at
least $\sim$10 times brighter than the upper limit on the X-ray
luminosity of 2MASS\,J0414, hence it would be easily detectable. Yet,
we do not detect 2MASS\,J0414 in X-rays during the $U$-band maximum of
luminosity. Moreover, the luminosity increase lasts for at least 6\,h
and exhibits a slow continuous rise. In contrast, the $U$-band flares
observed from Proxima~Cen are impulsive, and the duration of the
longest flare is shorter than one hour \citep{guedel04}. We thus
conclude that the $U$-band luminosity rise is unlikely to result from
a magnetic flare.

Emission line variability occurring on timescales of hours has
been reported for young BDs by \citet{scholz06}. The
line variability is interpreted as non-steady accretion onto the substellar
surface with the accretion rate varying by one order of magnitude over a few
hours. The $U$-band luminosity rise observed for 2MASS\,J0414 could
similarly result from an accretion burst onto the central object. 

Assuming a purely radial free-fall motion of the gas from the disk's truncation
radius towards the corotating BD, the 6\,h timescale of the $U$-band
rise would correspond to the travel time from a distance of 3.5 substellar
radii (i.e., 1.8\,R$_\odot$) above the substellar
surface\footnote{The
ballistic infall velocity from rest at the disk's truncation radius,
$r_{\rm m}$, is: $V(r)^2=2 G M_\star \,(1/r-1/r_{\rm m})$
\citep[e.g.,][]{hartmann94}. Therefore, assuming a purely radial
infall, the travel time from the disk's truncation radius to the
substellar surface is: $t=r_{\rm
m}^{3/2}/\!\sqrt{2GM_\star}\times(\sin{2\theta}-2\theta+\pi)/2$, where
$\theta=\arccos{(\!\sqrt{1-R_\star/r_{\rm m}})}$.}; 
the terminal velocity would be about 200\,km\,s$^{-1}$. The Keplerian
period at the disk's truncation radius is then about 35\,h, which is
in the range of rotational periods observed for BDs
\citep[4--110\,h][]{scholz05b,zapatero06}.
The $U$-band increase of $\sim$1.1\,mag would then translate into a
mass accretion rate enhanced by a factor of nearly 3 compared to the
quiescent level, assuming that the $U$-band luminosity is directly
related to the accretion luminosity as observed for the T~Tauri stars \citep{gullbring98}.

As an alternative to the accretion burst, the 2MASS\,J0414 light curve
may simply reflect the rotational modulation of the object's luminosity by a hot
accretion shock located on the substellar surface. Evidence for
rotational modulation of the H$\alpha$ line profile by the accretion
flow has previously been reported in another young BD
\citep{scholz05a}. A hot accretion spot would result in the periodic
modulation of the light curve on a timescale of the BD rotational
period. 
The 6\,h rise time of 2MASS\,J0414's
$U$-band light curve suggests a rotational period of the order of 12\,h,
and is thus consistent with rotational modulation.  Assuming a
black-body temperature of 8000\,K for the accretion shock, the hot
spot would cover about 0.3\% of the substellar surface
\citep[e.g.,][]{bouvier93}. This value is similar to the size of
accretion shocks on classical T~Tauri stars.
The two subsequent $U$-band ground observations, obtained 6.4\,h
apart, show no sign of variability, unlike the OM light curve on this
timescale. This indicates that 2MASS\,J0414 was back to its quiescent
level nearly one year after the OM observations. Accretion hot spots
have been reported to be variable features in T~Tauri stars, and even
in one young BD \citep[see][ and references therein]{scholz05a}.

We conclude that at least two plausible sources of variability could
account for the observed $U$-band light curve of 2MASS\,J0414: an episodic
accretion event onto the central object or the rotational modulation of the
BD luminosity by a hot accretion spot. In both cases these
observations indicate variable mass accretion rate in this BD.

Our $U$-band survey of 11 BDs (of which 5 are actively
accreting and 4 are not accreting) indicates 2MASS\,J0414 as
the brightest source,
displaying variability related to accretion from
a circumsubstellar disk. 2MASS\,J0414 is then the best target of this
sample of 5 accreting BDs for ground follow-up observations to
investigate accretion in substellar objects, for example
with simultaneous $U$-band and spectroscopic ground monitoring with a
one-hour sampling.

\begin{acknowledgements}
We thank the International Space Science Institute (ISSI) in Bern for
significant financial support of the project team, Arlin Crotts for
organizing and obtaining photometry observations with the MDM on our
behalf, and the anonymous referee for useful comments. \xmm~is an ESA
science mission with instruments and contributions directly funded by
ESA Member States and the USA (NASA). X-ray astronomy research at PSI
has been supported by the Swiss National Science Foundation (grants
20-66875.01 and 20-109255/1). M.A.\ acknowledges support from NASA
grant NNG05GF92G.
\end{acknowledgements}

\appendix

\section{Formulae for the OM aperture photometry and associated errors}
\label{appendix}

We give here the formulae for the canonical 6\arcsec-radius
aperture that we used in our {\tt IDL} procedures for the OM aperture
photometry and associated errors.

The relation between the count rate of incident photons, $C_{\rm 0}$,
and the measure count rate on the detector, $C_{\rm det}$, is:
\begin{equation}
\label{eq:a1}
C_\mathrm{0}=F^\mathrm{\, theo}_\mathrm{ cl}(x,T)\times
G^\mathrm{ \,emp}_\mathrm{ cl}(x) \times f_\mathrm{ deadtime}(T) \times
f_\mathrm{tds}\,, 
\end{equation}
where $F^{\rm \,theo}_{\rm cl}$ is the theoretical formula for
coincidence-loss, $G^{\rm \,emp}_{\rm cl}$ is the
empirical correction of $F^{\rm\, theo}_{\rm cl}$ from
calibration observations, $f_{\rm deadtime}$ is the deadtime correction, and
$f_{\rm tds}$ is the temporal degradation sensitivity correction.  
$T$ is the frametime, i.e., the sum of the frame transfer time ($T_{\rm
ft}=0.1740$\,ms), needed to transfer the charges accumulated on the
continuously exposed CCD during an integration time into the storage
area, and the time needed to read out all the pixels in the storage
area. The frametime depends on the size and position of the exposure
window; here it is in units of seconds, and $x \equiv C_{\rm det}
\times T $. Note that using an aperture radius different than the
6\arcsec~value used in the calibration of the coincidence-loss would
add a correction factor in Eq.~\ref{eq:a1} using the PSF calibration.

The theoretical correction of the coincidence-loss
\citep[see][]{ehle05} is given by: 
\begin{equation}
F^{\rm\, theo}_{\rm cl}(x,T)= \frac{\ln (1-x)}{T_{\rm ft}-T}
\, .
\end{equation}

The empirical correction of this theoretical formula is:
\begin{equation}
G^{\rm \,emp}_{\rm cl}(x)=1+\Sigma_{i=1}^4 \, a_i \, x^{\,i} \, ,
\end{equation}
where the polynomial coefficients are provided by the extension COINCIDENCE of
the current calibration file OM\_PHOTTONAT\_0003.CCF. 

The only time when source photons are not properly recorded by
the detector is during charge transfer; therefore the deadtime is
equal to the frame transfer time, and the deadtime fraction is $d \equiv
T_{\rm ft}/T$. Consequently, the deadtime correction is: $f_{\rm
deadtime}(T) \equiv 1/(1-d)=T/(T-T_{\rm ft})$. 

The temporal degradation sensitivity correction, $f_\mathrm{tds}$, is
available in {\tt SAS} version 6.5 in the fits header of the observation
source list (*OM*SWSRLI*).

The count rate error of incident photons, $\Delta C_{\rm 0}$, is
computed by Gaussian propagation in Eq.~\ref{eq:a1} of the Poissonian
count rate error on the detector, $\Delta C_{\rm det}=\sqrt{C_{\rm
det}/T_{\rm exp}}$, where $T_{\rm exp}$ is the exposure time, which
leads to:
\begin{eqnarray}
\frac{\Delta C_{\rm 0}}{\Delta C_{\rm det}} & = & \frac{f_\mathrm{tds} \, T^2}{(T-T_{\rm
ft})^2} \nonumber \\ 
& & \times \left\{ \frac{1 + \Sigma_{i=1}^4 \, a_i \, x^{\,i} }{1-x} -
\ln{(1-x)} \, \Sigma_{i=1}^4 \, i \, a_i \, x^{\,i-1} \right\} \,.
\end{eqnarray}

We calculate the error associated with an object count rate measured
with aperture photometry as the quadratic sum of the count rate error of
incident sky+object photons and the count rate error of incident sky
photons determined with previous equations.

\bibliographystyle{aa}
\bibliography{biblio}

\end{document}